\documentstyle[12pt,eepic,epic]{article}
\newlength{\minitwocolumn}
\title{Extracting visible $s-$channel Higgs in the lepton scattering}
\author{\vspace{1cm}\\
        {\bf L. T. Handoko} \thanks{On leave from P3FT-LIPI, Indonesia. 
        E-mail address : handoko@theo.phys.sci.hiroshima-u.ac.jp} \\
        Department of Physics, Hiroshima University \\
        1-3-1 Kagamiyama, Higashi Hiroshima - 739, Japan\\
        \vspace{3mm}\\}
\date{}

\begin{document}
\setlength{\baselineskip}{24pt}

\maketitle
\begin{picture}(0,0)
       \put(325,260){HUPD-9614}
       \put(325,245){August 1996}
\end{picture}
\vspace{-24pt}

\setlength{\baselineskip}{8mm}
\thispagestyle{empty}
\renewcommand{\thesubsection}{\Roman{subsection}}

\begin{abstract}
	The contributions of $s-$channel Higgs in 
	$l^+ \, l^- \rightarrow q \, \bar{q}$ processes in the general 
	lepton collider ($l^+ \, l^-$ collider) within Standard 
	Model is studied. A new idea to extract the contribution by 
	using the data from both electron and future heavy-lepton 
	colliders at same center-of-mass energy is proposed. 
	Deviations due to the $s-$channel Higgs contributions are analysed 
	and discussed for the heavy-quark final states by using the total 
	cross-section, forward-backward asymmetry and its ratio. It is 
	shown that significant deviations are expected for the top quark 
	final state.
\end{abstract}

\clearpage

\section{\bf Introduction}
\label{sec:introduction}

Although the development of the lepton colliders beyond the electron 
collider are at a very early stage, its promise for physics is clear. 
However, now the possibility of constructing a muon collider is coming 
into the limelight. The efforts to construct the {\it heavy-lepton} 
collider are especially motivated by the limitations of the electron 
collider to achieve high center-of-mass energy\cite{electron, muon}. 
High center-of-mass 
energy is necessary to study the behaviour of Higgs particle in detail 
as well as open a window of new physics \cite{gunion}. In this letter, 
the Higgs particle is studied by using the total cross-section (CS), 
forwad-backward (FB) asymmetry and its ratio ($R$) of two jets productions 
in the lepton scattering, $l^+ \, l^- \rightarrow q \, \bar{q}$ 
within tree-level Standard Model (SM). 
The reason is clear, because these quantities are complementary, 
i.e. the terms which contribute to the total CS will be nothing in the 
FB asymmetry and vice-versa. On the other hand, the ratio of FB asymmetry 
and total CS gives a clear and may be experimentally accessible quantity, 
because any uncertainties in both theoretical and experimental sides 
are reduced. 
Next point of this paper is, a trial to make the $s-$channel Higgs 
contributions to be more visible by using the data from both electron 
collider and future heavy-lepton colliders together. These points are 
the originality of this paper. 

In general, the FB asymmetry and total CS for the initial state $l$ 
is obtained by integrating the CS ($\sigma$) with respect to the 
angular variable $z$ ($\equiv \cos \theta$) and defining, 
\begin{equation}
	{A_{\rm FB}}^l \equiv {\sigma_F}^l - {\sigma_B}^l \; ,
	\label{eq:fba}
\end{equation}
for the FB asymmetry and 
\begin{equation}
	{\sigma_T}^l \equiv {\sigma_F}^l + {\sigma_B}^l \; ,
	\label{eq:tcs}
\end{equation}
for the total CS. Here, the forward and backward scattered CS are given as, 
\begin{eqnarray}
	\sigma_F & \equiv & \int^1_0 \frac{{\rm d} \sigma}{{\rm d} z}
		\, {\rm d} z \; , 
	\label{eq:sf} \\
	\sigma_B & \equiv & \int^0_{-1} \frac{{\rm d} \sigma}{{\rm d} z}
		\, {\rm d} z \; .
	\label{eq:sb}
\end{eqnarray}
These quantities lead to the ratio $R$ to be defined as follows 
\begin{equation}
	R^l \equiv \frac{{A_{\rm FB}}^l}{{\sigma_T}^l} \; .
	\label{eq:r}
\end{equation}
In the next section, the calculation of this FB asymmetry, total CS
and also how the significant contributions of the $s-$channel Higgs can 
be extracted will be shown. 

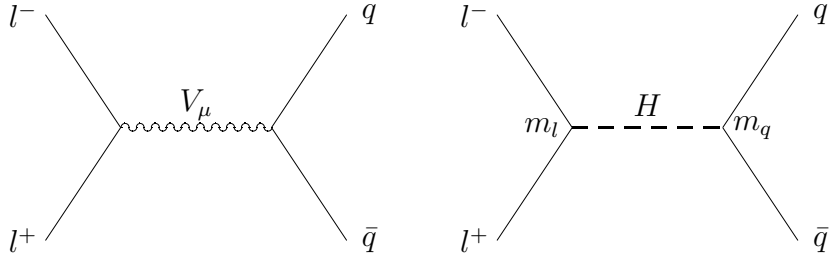
\begin{figure}[t]
	\unitlength 1mm
        \begin{center}
	\begin{picture}(100,30)
		\put(0,30){\line(2,-3){10}}
		\put(0,0){\line(2,3){10}}
		\put(40,0){\line(-2,3){10}}
		\put(40,30){\line(-2,-3){10}}
		\put(60,30){\line(2,-3){10}}
		\put(60,0){\line(2,3){10}}
		\put(100,0){\line(-2,3){10}}
		\put(100,30){\line(-2,-3){10}}
		\multiput(10.5,15)(2,0){10}{\oval(1,1)[t]}
		\multiput(11.5,15)(2,0){10}{\oval(1,1)[b]}
		\multiput(70,15)(3.5,0){6}{\line(1,0){2}}
		\put(-3,30){\makebox(0,0){$l^-$}}
		\put(-3,0){\makebox(0,0){$l^+$}}
		\put(57,30){\makebox(0,0){$l^-$}}
		\put(57,0){\makebox(0,0){$l^+$}}
		\put(103,30){\makebox(0,0){$q$}}
		\put(103,0){\makebox(0,0){$\bar{q}$}}
		\put(43,30){\makebox(0,0){$q$}}
		\put(43,0){\makebox(0,0){$\bar{q}$}}
		\put(20,18){\makebox(0,0){$V_\mu$}}
		\put(80,18){\makebox(0,0){$H$}}
		\put(66,15){\makebox(0,0){$m_l$}}
		\put(94,15){\makebox(0,0){$m_q$}}
	\end{picture}
        \end{center}
        \caption{The tree-level diagrams in the 
		$l^+ \, l^- \rightarrow q \, \bar{q}$ process 
		within the SM in the unitary gauge.}
        \label{fig:diagrams}
\end{figure}

\section{\bf Calculation}
\label{sec:calculation}

Within tree-level SM, $l^+ \, l^- \rightarrow q \, \bar{q}$ process is 
realized in the vector bosons ($Z$ and photon) and also scalar Higgs 
mediated diagrams as depicted in Fig. \ref{fig:diagrams}. The related 
interactions are expressed as, 
\begin{eqnarray}
	{\cal L}_V & = & g \, \bar{f} \, \gamma^\mu \, \left( 
		{g_{VL}}^f \, L + {g_{VR}}^f \, R \right) \, f \, V_\mu \; , 
	\label{eq:lv} \\
	{\cal L}_H & = & -g \, \frac{m_f}{2 \, M_W} \, \bar{f} \, f \, H \; .
	\label{eq:lh}
\end{eqnarray}
Here, $f$ denotes fermions, $L$ and $R$ are the chiralities and $V = Z, A$. 
The couplings for the vector bosons are given as, 
\begin{eqnarray}
	{g_A}^f & \equiv & {g_{AL}}^f = {g_{AR}}^f \equiv 
		Q_f \, \sin \theta_W \; , 
	\label{eq:ga} \\ 
	{g_{ZL}}^f & \equiv & \frac{1}{2 \, \cos \theta_W} \left( 
		\pm 1 - 2 \, Q_f \, \sin^2 \theta_W \right) \; , 
	\label{eq:gzl} \\
	{g_{ZR}}^f & \equiv & - \frac{\sin^2 \theta_W}{\cos \theta_W} \, 
		Q_f \; ,
	\label{eq:gzr}
\end{eqnarray}
with $Q_l = -1$, $Q_u = 2/3$, $Q_d = -1/3$ and $\theta_W$ is the Weinberg 
angle respectively. The sign $\pm$ means $+$ for up-quarks, while $-$ for 
down-quarks and leptons.

Remind that ${{\rm d}\sigma}/{{\rm d}z}$ is proportioned to 
$\left| M \right|^2$. So, in order to accomplish the FB asymmetry and 
total CS in Eqs. (\ref{eq:fba}) and (\ref{eq:tcs}), one has to compute 
the square amplitudes of the diagrams in Fig. \ref{fig:diagrams}, that is 
\begin{equation}
	\left| M \right|^2 = 24 \, {G_F}^2 \left[
		\left| M_Z \right|^2 + \left| M_A \right|^2 + 
		\left| M_H \right|^2 + 2 \, {\rm Re} \left( 
		M_Z \, {M_A}^\dagger + M_Z \, {M_H}^\dagger + 
		M_A \, {M_H}^\dagger \right) \right] \; , 
	\label{eq:msquare}
\end{equation}
with $M_i$ denotes the amplitude of each gauge bosons and scalar Higgs. 
Note that color and spin averaged factors have been included.
Although the calculation is quite trivial, it is better to present the 
analytic results for comparison and further analysis. Then, each term 
in Eq. (\ref{eq:msquare}) is given as, 
\begin{eqnarray}
	\left| M_Z \right|^2 & = & \frac{{M_W}^4}{
		\left( s - {M_Z}^2 \right)^2 + {M_Z}^2 \, {\Gamma_Z}^2} \, 
		\nonumber \\
	& &	\times \left[ \left( ({g_{ZL}}^l)^2 + 
			({g_{ZR}}^l)^2 \right) \, 
		\left( ({g_{ZL}}^q)^2 + ({g_{ZR}}^q)^2 \right) \, 
		\right. \nonumber \\
	& & \left. \; \; \; \; \; \; \; \; \; \; \; \; \; \; \; \; \; \; 
		\times \left( s^2 + {u(s)}^2 \, \cos^2 \theta - 
			\frac{2}{{M_Z}^2} \, s^3 \right) 
		\right. \nonumber \\
	& & \left. \; \; \; \; \; \; 
		- 2 \, \left( ({g_{ZL}}^l)^2 - ({g_{ZR}}^l)^2 \right) \,
		\left( ({g_{ZL}}^q)^2 - ({g_{ZR}}^q)^2 \right) \, 
		\right. \nonumber \\
	& & \left. \; \; \; \; \; \; \; \; \; \; \; \; \; \; \; \; \; \; 
		\times u(s) \, s \, \cos \theta \, \left( 
			1 - \frac{4}{{M_Z}^2} \, s \right)
		\right. \nonumber \\
	& & \left. \; \; \; \; \; \; 
		+ 8 \, {m_q}^2 \, {g_{ZL}}^q \, {g_{ZR}}^q \, 
		\left( ({g_{ZL}}^l)^2 + ({g_{ZR}}^l)^2 \right) \, 
		\left( s - 2 \, {m_l}^2 + \frac{s^2}{{M_Z}^2} \right)	
		\right. \nonumber \\
	& & \left. \; \; \; \; \; \; 
		+ 8 \, {m_l}^2 \, {g_{ZL}}^l \, {g_{ZR}}^l \, 
		\left( ({g_{ZL}}^q)^2 + ({g_{ZR}}^q)^2 \right) \, 
		\left( s - 2 \, {m_q}^2 + \frac{s^2}{{M_Z}^2} \right)	
		\right. \nonumber \\
	& & \left. \; \; \; \; \; \; 
		- \frac{32}{{M_Z}^2} \, {m_q}^2 \, {m_l}^2 \, 
		{g_{ZL}}^l \, {g_{ZR}}^l \, {g_{ZL}}^q \, {g_{ZR}}^q \, s 
		\right. \nonumber \\
	& & \left. \; \; \; \; \; \; 
		+ \frac{4}{{M_Z}^4} \, 
			\left( {g_{ZL}}^l - {g_{ZR}}^l \right)^2 
			\, \left( {g_{ZL}}^q - {g_{ZR}}^q \right)^2 \, 
			{m_q}^2 \, {m_l}^2 \, s^2 	
		\right] \; , 
	\label{eq:mzsquare}\\
	\left| M_A \right|^2 & = & 
		\frac{{M_W}^4}{s^2} \, 
		\left( {g_A}^l \, {g_A}^q \right)^2 \, 
		\left[ s^2 + {u(s)}^2 \, \cos^2 \theta + 
		4 \, s \, \left( {m_l}^2 + {m_q}^2 \right) \right] \; , 
	\label{eq:masquare}\\
	\left| M_H \right|^2 & = & \frac{1}{16} \, 
		\frac{{m_l}^2 \, {m_q}^2}{
		\left( s - {M_H}^2 \right)^2 + {M_H}^2 \, {\Gamma_H}^2} \, 
		{u(s)}^2 \; , 
	\label{eq:mhsquare}\\
	{\rm Re} \left( M_Z \, {M_A}^\dagger \right) & = & 
		\frac{{M_W}^4}{s} \, \frac{s - {M_Z}^2}{
		\left( s - {M_Z}^2 \right)^2 + {M_Z}^2 \, {\Gamma_Z}^2} \, 
	 	{g_A}^l \, {g_A}^q \, 
		\nonumber \\
	& & 	\times \left\{ \left( {g_{ZL}}^l + {g_{ZR}}^l \right) \, 
		\left( {g_{ZL}}^q + {g_{ZR}}^q  \right) \, 
		\right. \nonumber \\
	& &	\left. \; \; \; \; \; \; \; \; \; \; \; \; 
		\times \left[ s^2 + {u(s)}^2 \, \cos^2 \theta + 
		4 \, s \, \left( {m_l}^2 + {m_q}^2 \right) \right] 
		\right. \nonumber \\
	& & \left. \; \; \; \; \; \; 
		- 2 \, 	\left( {g_{ZL}}^l - {g_{ZR}}^l \right) \, 
		\left( {g_{ZL}}^q - {g_{ZR}}^q  \right) \, 
		u(s) \, s \, \cos \theta \right\} \; , 
	\label{eq:mzma}\\
	{\rm Re} \left( M_Z \, {M_H}^\dagger \right) & = & 
		\frac{\left( s - {M_H}^2 \right) \left( s - {M_Z}^2 \right)
			+ M_H \, M_Z \, \Gamma_H \, \Gamma_Z}{
		\left[ 
		\left( s - {M_H}^2 \right)^2 + {M_H}^2 \, {\Gamma_H}^2 \right]
		\left[ 
		\left( s - {M_Z}^2 \right)^2 + {M_Z}^2 \, {\Gamma_Z}^2 \right]}
		\nonumber \\
	& & 	\times 
		{M_W}^2 \, {m_l}^2 \, {m_q}^2 \, 
		\left( {g_{ZL}}^l + {g_{ZR}}^l \right) \, 
		\left( {g_{ZL}}^q + {g_{ZR}}^q  \right) \, 
		u(s) \, \cos \theta \; , 
	\label{eq:mzmh}\\
	{\rm Re} \left( M_A \, {M_H}^\dagger \right) & = & 
		{M_W}^2 \, \frac{{m_l}^2 \, {m_q}^2}{s} \, 
		\frac{s - {M_H}^2}{
		\left( s - {M_H}^2 \right)^2 + {M_H}^2 \, {\Gamma_H}^2} \, 
		{g_A}^l \, {g_A}^q \, u(s) \, \cos \theta \; ,  
	\label{eq:mamh}
\end{eqnarray}
where $\sqrt{s}$ is center-of-mass energy and $u(s) \equiv \sqrt{
\left( s - 4 \, {m_l}^2 \right) \left( s - 4 \, {m_q}^2  \right)}$. 
$u(s)$ also expresses the boundary condition for the physical region in 
the process, i.e. $s \geq 4 \, {m_q}^2$ and $s \geq 4 \, {m_l}^2$ 
as well. From Eqs. (\ref{eq:mhsquare}), (\ref{eq:mzmh}) and (\ref{eq:mamh}), 
it is clear that significant deviations due to the $s-$channel Higgs 
contributions would be coming out for the heavy-quark final states. 
Hence, further discussion will be emphasized only on the top and bottom 
quark final states.

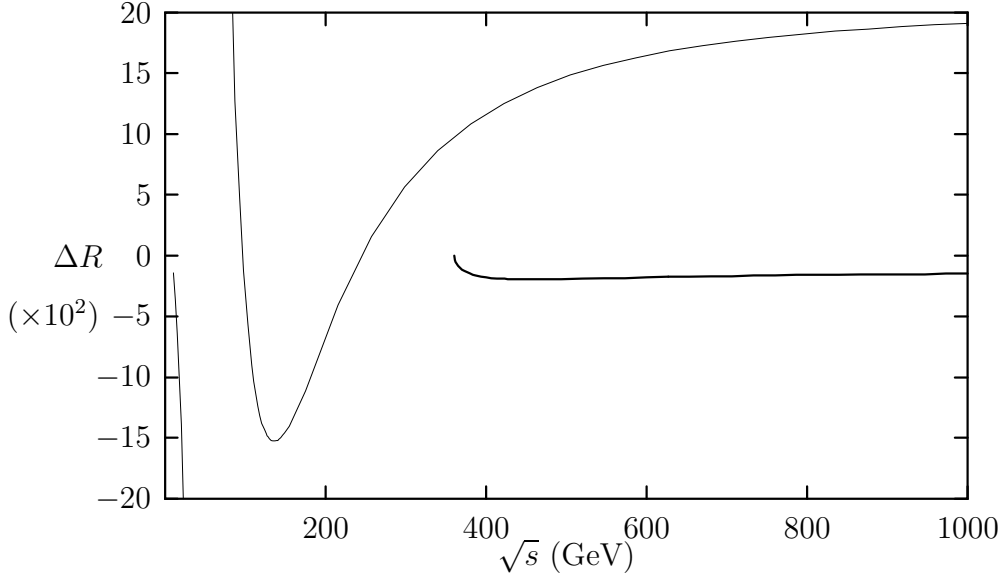
\begin{figure}[t]
        \begin{center}
\setlength{\unitlength}{0.240900pt}
\begin{picture}(1500,900)(0,0)
\thicklines \path(176,113)(196,113)
\thicklines \path(1436,113)(1416,113)
\put(154,113){\makebox(0,0)[r]{$-20$}}
\thicklines \path(176,209)(196,209)
\thicklines \path(1436,209)(1416,209)
\put(154,209){\makebox(0,0)[r]{$-15$}}
\thicklines \path(176,304)(196,304)
\thicklines \path(1436,304)(1416,304)
\put(154,304){\makebox(0,0)[r]{$-10$}}
\thicklines \path(176,400)(196,400)
\thicklines \path(1436,400)(1416,400)
\put(154,400){\makebox(0,0)[r]{$-5$}}
\thicklines \path(176,495)(196,495)
\thicklines \path(1436,495)(1416,495)
\put(154,495){\makebox(0,0)[r]{$0$}}
\thicklines \path(176,591)(196,591)
\thicklines \path(1436,591)(1416,591)
\put(154,591){\makebox(0,0)[r]{$5$}}
\thicklines \path(176,686)(196,686)
\thicklines \path(1436,686)(1416,686)
\put(154,686){\makebox(0,0)[r]{$10$}}
\thicklines \path(176,782)(196,782)
\thicklines \path(1436,782)(1416,782)
\put(154,782){\makebox(0,0)[r]{$15$}}
\thicklines \path(176,877)(196,877)
\thicklines \path(1436,877)(1416,877)
\put(154,877){\makebox(0,0)[r]{$20$}}
\thicklines \path(428,113)(428,133)
\thicklines \path(428,877)(428,857)
\put(428,68){\makebox(0,0){$200$}}
\thicklines \path(680,113)(680,133)
\thicklines \path(680,877)(680,857)
\put(680,68){\makebox(0,0){$400$}}
\thicklines \path(932,113)(932,133)
\thicklines \path(932,877)(932,857)
\put(932,68){\makebox(0,0){$600$}}
\thicklines \path(1184,113)(1184,133)
\thicklines \path(1184,877)(1184,857)
\put(1184,68){\makebox(0,0){$800$}}
\thicklines \path(1436,113)(1436,133)
\thicklines \path(1436,877)(1436,857)
\put(1436,68){\makebox(0,0){$1000$}}
\thicklines \path(176,113)(1436,113)(1436,877)(176,877)(176,113)
\put(806,23){\makebox(0,0){$\sqrt{s}$ (GeV)}}
\put(0,495){\makebox(0,0)[l]{$\Delta R$}}
\put(0,400){\makebox(0,0){$(\times 10^2)$}}
\thinlines \path(189,468)(189,468)(192,426)(195,373)(202,230)(205,113)
\thinlines \path(282,877)(286,738)(293,586)(299,472)(306,387)(312,323)(315,298)(319,277)(322,259)(325,244)(328,232)(332,222)(335,215)(336,212)(338,210)(340,208)(341,206)(343,205)(345,204)(346,204)(348,204)(349,204)(351,205)(353,205)(354,206)(358,209)(361,213)(364,217)(371,227)(397,283)(448,418)(500,525)(552,603)(604,660)(656,702)(708,734)(760,759)(812,779)(864,794)(916,806)(968,817)(1020,825)(1072,832)(1124,838)(1176,843)(1228,848)(1280,851)(1332,855)(1384,858)(1436,860)
\thicklines \path(630,495)(630,495)(631,488)(632,485)(634,482)(636,479)(638,477)(642,473)(646,471)(655,467)(659,465)(663,464)(672,462)(680,461)(688,460)(697,459)(705,459)(709,459)(714,458)(722,458)(726,458)(730,458)(735,458)(737,458)(739,458)(741,458)(742,458)(743,458)(744,458)(745,458)(746,458)(747,458)(748,458)(749,458)(750,458)(751,458)(752,458)(753,458)(755,458)(756,458)(760,458)(762,458)(764,458)(772,458)(781,458)(798,458)(831,459)(865,460)(898,460)(932,461)(966,462)
\thicklines \path(966,462)(999,462)(1033,463)(1066,463)(1100,464)(1134,464)(1167,465)(1201,465)(1234,465)(1268,466)(1302,466)(1335,466)(1369,466)(1402,467)(1436,467)
\end{picture}
        \end{center}
        \caption{$\Delta R$ as a function of center-of-mass energy for top 
		(thick line) and bottom (thin line) quark final states 
		including Higgs contributions in the electron vs muon 
		collider case.}
        \label{fig:tb}
\end{figure}

\begin{figure}[t]
        \begin{minipage}[t]{\minitwocolumn}
        \begin{center}
\setlength{\unitlength}{0.240900pt}
\begin{picture}(1049,900)(0,0)
\thicklines \path(176,113)(196,113)
\thicklines \path(985,113)(965,113)
\put(154,113){\makebox(0,0)[r]{$0$}}
\thicklines \path(176,266)(196,266)
\thicklines \path(985,266)(965,266)
\put(154,266){\makebox(0,0)[r]{$2$}}
\thicklines \path(176,419)(196,419)
\thicklines \path(985,419)(965,419)
\put(154,419){\makebox(0,0)[r]{$4$}}
\thicklines \path(176,571)(196,571)
\thicklines \path(985,571)(965,571)
\put(154,571){\makebox(0,0)[r]{$6$}}
\thicklines \path(176,724)(196,724)
\thicklines \path(985,724)(965,724)
\put(154,724){\makebox(0,0)[r]{$8$}}
\thicklines \path(176,877)(196,877)
\thicklines \path(985,877)(965,877)
\put(154,877){\makebox(0,0)[r]{$10$}}
\thicklines \path(338,113)(338,133)
\thicklines \path(338,877)(338,857)
\put(338,68){\makebox(0,0){$200$}}
\thicklines \path(500,113)(500,133)
\thicklines \path(500,877)(500,857)
\put(500,68){\makebox(0,0){$400$}}
\thicklines \path(661,113)(661,133)
\thicklines \path(661,877)(661,857)
\put(661,68){\makebox(0,0){$600$}}
\thicklines \path(823,113)(823,133)
\thicklines \path(823,877)(823,857)
\put(823,68){\makebox(0,0){$800$}}
\thicklines \path(985,113)(985,133)
\thicklines \path(985,877)(985,857)
\put(985,68){\makebox(0,0){$1000$}}
\thicklines \path(176,113)(985,113)(985,877)(176,877)(176,113)
\put(580,23){\makebox(0,0){$\sqrt{s}$ (GeV)}}
\put(580,930){\makebox(0,0){$\Delta \sigma_T \; (\times 10^{14})$}}
\thinlines \path(467,844)(467,844)(489,737)(510,653)(532,585)(554,529)(575,483)(597,444)(618,411)(640,383)(661,358)(683,337)(705,319)(726,302)(748,288)(769,275)(791,264)(812,254)(834,244)(856,236)(877,228)(899,221)(920,215)(942,209)(963,204)(985,199)
\thicklines \path(467,844)(467,844)(489,775)(510,707)(532,646)(554,592)(575,545)(597,504)(618,468)(640,437)(661,409)(683,385)(705,364)(726,345)(748,328)(769,312)(791,299)(812,287)(834,275)(856,265)(877,256)(899,248)(920,240)(942,233)(963,227)(985,221)
\end{picture}
        \end{center}
        \end{minipage}
        \hfill
        \begin{minipage}[t]{\minitwocolumn}
        \begin{center}
\setlength{\unitlength}{0.240900pt}
\begin{picture}(1049,900)(0,0)
\thicklines \path(176,164)(196,164)
\thicklines \path(985,164)(965,164)
\put(154,164){\makebox(0,0)[r]{$-14$}}
\thicklines \path(176,266)(196,266)
\thicklines \path(985,266)(965,266)
\put(154,266){\makebox(0,0)[r]{$-12$}}
\thicklines \path(176,368)(196,368)
\thicklines \path(985,368)(965,368)
\put(154,368){\makebox(0,0)[r]{$-10$}}
\thicklines \path(176,470)(196,470)
\thicklines \path(985,470)(965,470)
\put(154,470){\makebox(0,0)[r]{$-8$}}
\thicklines \path(176,571)(196,571)
\thicklines \path(985,571)(965,571)
\put(154,571){\makebox(0,0)[r]{$-6$}}
\thicklines \path(176,673)(196,673)
\thicklines \path(985,673)(965,673)
\put(154,673){\makebox(0,0)[r]{$-4$}}
\thicklines \path(176,775)(196,775)
\thicklines \path(985,775)(965,775)
\put(154,775){\makebox(0,0)[r]{$-2$}}
\thicklines \path(176,877)(196,877)
\thicklines \path(985,877)(965,877)
\put(154,877){\makebox(0,0)[r]{$0$}}
\thicklines \path(176,113)(176,133)
\thicklines \path(176,877)(176,857)
\put(176,68){\makebox(0,0){$0$}}
\thicklines \path(338,113)(338,133)
\thicklines \path(338,877)(338,857)
\put(338,68){\makebox(0,0){$200$}}
\thicklines \path(500,113)(500,133)
\thicklines \path(500,877)(500,857)
\put(500,68){\makebox(0,0){$400$}}
\thicklines \path(661,113)(661,133)
\thicklines \path(661,877)(661,857)
\put(661,68){\makebox(0,0){$600$}}
\thicklines \path(823,113)(823,133)
\thicklines \path(823,877)(823,857)
\put(823,68){\makebox(0,0){$800$}}
\thicklines \path(985,113)(985,133)
\thicklines \path(985,877)(985,857)
\put(985,68){\makebox(0,0){$1000$}}
\thicklines \path(176,113)(985,113)(985,877)(176,877)(176,113)
\put(580,23){\makebox(0,0){$\sqrt{s}$ (GeV)}}
\put(580,930){\makebox(0,0){$\Delta A_{\rm FB} \; (\times 10^{16})$}}
\thinlines \path(467,877)(467,877)(468,850)(469,839)(470,823)(471,811)(473,801)(478,772)(483,752)(489,736)(500,714)(505,705)(510,698)(516,693)(521,688)(527,685)(532,682)(537,680)(543,678)(545,677)(548,677)(551,676)(552,676)(554,676)(555,676)(556,676)(558,676)(558,676)(559,676)(560,676)(560,676)(561,676)(562,676)(562,676)(563,676)(564,676)(564,676)(566,676)(567,676)(570,676)(572,677)(575,677)(586,679)(597,681)(618,688)(640,695)(661,704)(683,712)(705,720)(726,728)(748,736)
\thinlines \path(748,736)(769,743)(791,750)(812,756)(834,763)(856,768)(877,774)(899,779)(920,783)(942,788)(963,792)(985,796)
\thicklines \path(467,877)(467,877)(468,702)(469,633)(470,539)(471,472)(473,420)(475,341)(478,285)(481,243)(483,211)(486,186)(489,168)(492,155)(493,149)(494,145)(496,142)(497,139)(498,137)(500,135)(500,134)(501,134)(502,134)(502,133)(503,133)(504,133)(504,133)(505,134)(506,134)(506,134)(508,135)(509,137)(510,138)(516,147)(521,159)(532,189)(554,258)(575,325)(597,386)(618,440)(640,487)(661,527)(683,562)(705,592)(726,618)(748,640)(769,660)(791,678)(812,693)(834,707)(856,719)
\thicklines \path(856,719)(877,730)(899,740)(920,749)(942,757)(963,765)(985,771)
\end{picture}
        \end{center}
        \end{minipage}
        \caption{$\Delta \sigma_T$ (left) and $\Delta A_{\rm FB}$ (right) as a 
		function of center-of-mass energy for top quark final state 
		including (thick line) and excluding (thin line) Higgs 
		contributions in the electron vs muon collider case.}
        \label{fig:tcsfba}
\end{figure}
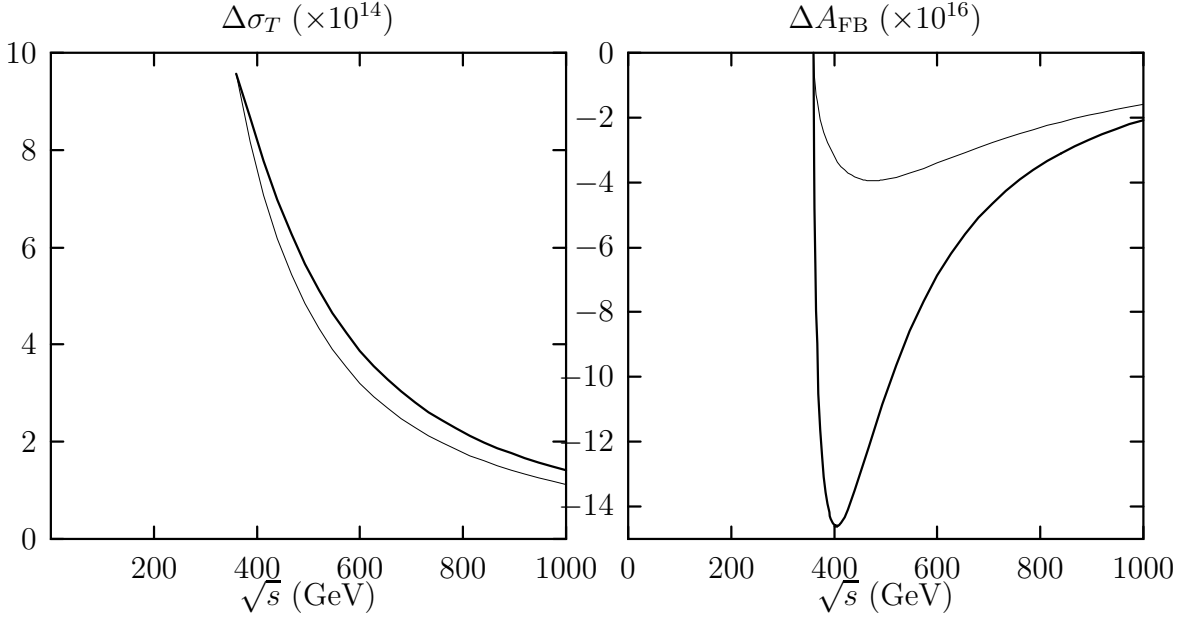

\begin{figure}[t]
        \begin{minipage}[t]{\minitwocolumn}
        \begin{center}
\setlength{\unitlength}{0.240900pt}
\begin{picture}(1049,900)(0,0)
\thicklines \path(176,113)(196,113)
\thicklines \path(985,113)(965,113)
\put(154,113){\makebox(0,0)[r]{$-2$}}
\thicklines \path(176,189)(196,189)
\thicklines \path(985,189)(965,189)
\thicklines \path(176,266)(196,266)
\thicklines \path(985,266)(965,266)
\thicklines \path(176,342)(196,342)
\thicklines \path(985,342)(965,342)
\thicklines \path(176,419)(196,419)
\thicklines \path(985,419)(965,419)
\thicklines \path(176,495)(196,495)
\thicklines \path(985,495)(965,495)
\put(154,495){\makebox(0,0)[r]{$-1$}}
\thicklines \path(176,571)(196,571)
\thicklines \path(985,571)(965,571)
\thicklines \path(176,648)(196,648)
\thicklines \path(985,648)(965,648)
\thicklines \path(176,724)(196,724)
\thicklines \path(985,724)(965,724)
\thicklines \path(176,801)(196,801)
\thicklines \path(985,801)(965,801)
\thicklines \path(176,877)(196,877)
\thicklines \path(985,877)(965,877)
\put(154,877){\makebox(0,0)[r]{$0$}}
\thicklines \path(176,113)(176,133)
\thicklines \path(176,877)(176,857)
\put(176,68){\makebox(0,0){$0$}}
\thicklines \path(338,113)(338,133)
\thicklines \path(338,877)(338,857)
\put(338,68){\makebox(0,0){$200$}}
\thicklines \path(500,113)(500,133)
\thicklines \path(500,877)(500,857)
\put(500,68){\makebox(0,0){$400$}}
\thicklines \path(661,113)(661,133)
\thicklines \path(661,877)(661,857)
\put(661,68){\makebox(0,0){$600$}}
\thicklines \path(823,113)(823,133)
\thicklines \path(823,877)(823,857)
\put(823,68){\makebox(0,0){$800$}}
\thicklines \path(985,113)(985,133)
\thicklines \path(985,877)(985,857)
\put(985,68){\makebox(0,0){$1000$}}
\thicklines \path(176,113)(985,113)(985,877)(176,877)(176,113)
\put(900,930){\makebox(0,0)[l]{$\Delta R \; (\times 10^2)$}}
\put(580,23){\makebox(0,0){$\sqrt{s}$ (GeV)}}
\thinlines \path(467,877)(467,877)(468,856)(469,847)(470,834)(473,815)(478,788)(489,748)(510,687)(532,640)(554,600)(575,567)(597,538)(618,513)(640,491)(661,472)(683,455)(705,440)(726,427)(748,415)(769,404)(791,394)(812,386)(834,378)(856,371)(877,364)(899,358)(920,352)(942,347)(963,343)(985,338)
\thicklines \path(467,877)(467,877)(468,740)(469,684)(470,609)(471,554)(473,510)(475,442)(478,390)(483,315)(486,287)(489,263)(494,226)(500,199)(505,179)(510,164)(516,154)(518,150)(521,146)(527,141)(529,139)(532,137)(535,136)(536,136)(537,135)(539,135)(539,135)(540,135)(541,135)(541,135)(542,135)(543,135)(543,135)(544,135)(545,135)(545,135)(546,135)(547,135)(547,135)(548,135)(551,135)(552,135)(554,136)(559,137)(564,139)(575,144)(597,157)(618,172)(640,186)(661,200)(683,212)
\thicklines \path(683,212)(705,224)(726,235)(748,245)(769,254)(791,263)(812,271)(834,278)(856,284)(877,290)(899,296)(920,301)(942,305)(963,310)(985,314)
\end{picture}
        \end{center}
        \end{minipage}
        \hfill
        \begin{minipage}[t]{\minitwocolumn}
        \begin{center}
\setlength{\unitlength}{0.240900pt}
\begin{picture}(1049,900)(0,0)
\thicklines \path(176,113)(196,113)
\thicklines \path(985,113)(965,113)
\put(154,113){\makebox(0,0)[r]{$-3$}}
\thicklines \path(176,240)(196,240)
\thicklines \path(985,240)(965,240)
\put(154,240){\makebox(0,0)[r]{$-2$}}
\thicklines \path(176,368)(196,368)
\thicklines \path(985,368)(965,368)
\put(154,368){\makebox(0,0)[r]{$-1$}}
\thicklines \path(176,495)(196,495)
\thicklines \path(985,495)(965,495)
\put(154,495){\makebox(0,0)[r]{$0$}}
\thicklines \path(176,622)(196,622)
\thicklines \path(985,622)(965,622)
\put(154,622){\makebox(0,0)[r]{$1$}}
\thicklines \path(176,750)(196,750)
\thicklines \path(985,750)(965,750)
\put(154,750){\makebox(0,0)[r]{$2$}}
\thicklines \path(176,877)(196,877)
\thicklines \path(985,877)(965,877)
\put(154,877){\makebox(0,0)[r]{$3$}}
\thicklines \path(176,113)(176,133)
\thicklines \path(176,877)(176,857)
\put(176,68){\makebox(0,0){$0$}}
\thicklines \path(338,113)(338,133)
\thicklines \path(338,877)(338,857)
\put(338,68){\makebox(0,0){$200$}}
\thicklines \path(500,113)(500,133)
\thicklines \path(500,877)(500,857)
\put(500,68){\makebox(0,0){$400$}}
\thicklines \path(661,113)(661,133)
\thicklines \path(661,877)(661,857)
\put(661,68){\makebox(0,0){$600$}}
\thicklines \path(823,113)(823,133)
\thicklines \path(823,877)(823,857)
\put(823,68){\makebox(0,0){$800$}}
\thicklines \path(985,113)(985,133)
\thicklines \path(985,877)(985,857)
\put(985,68){\makebox(0,0){$1000$}}
\thicklines \path(176,113)(985,113)(985,877)(176,877)(176,113)
\put(580,23){\makebox(0,0){$M_H$ (GeV)}}
\thicklines \path(216,305)(216,305)(232,303)(248,301)(264,297)(280,293)(297,288)(313,282)(329,274)(345,265)(361,254)(377,240)(409,203)(425,179)(441,153)(445,147)(449,142)(451,140)(453,138)(455,136)(456,136)(457,135)(458,135)(459,135)(460,135)(461,135)(462,135)(463,136)(464,136)(465,137)(466,139)(467,140)(469,145)(471,150)(473,158)(475,167)(477,178)(481,208)(485,250)(489,302)(497,439)(505,589)(509,656)(513,710)(515,732)(517,751)(519,767)(521,779)(523,788)(524,792)(525,795)
\thicklines \path(525,795)(526,797)(527,799)(528,800)(529,801)(530,802)(531,802)(532,801)(533,801)(534,800)(535,798)(537,795)(541,786)(545,775)(553,750)(569,701)(585,660)(601,627)(617,602)(633,581)(649,565)(665,551)(697,530)(729,515)(761,504)(793,495)(825,489)(857,483)(889,478)(921,474)(953,471)(985,468)
\end{picture}
        \end{center}
        \end{minipage}
        \caption{$\Delta R$ as a function of center-of-mass energy (left) 
		including (thick line) and excluding (thin line) 
		Higgs contributions, and Higgs mass (right) for top quark 
		final state in the electron vs muon collider case.}
        \label{fig:ratio}
\end{figure}
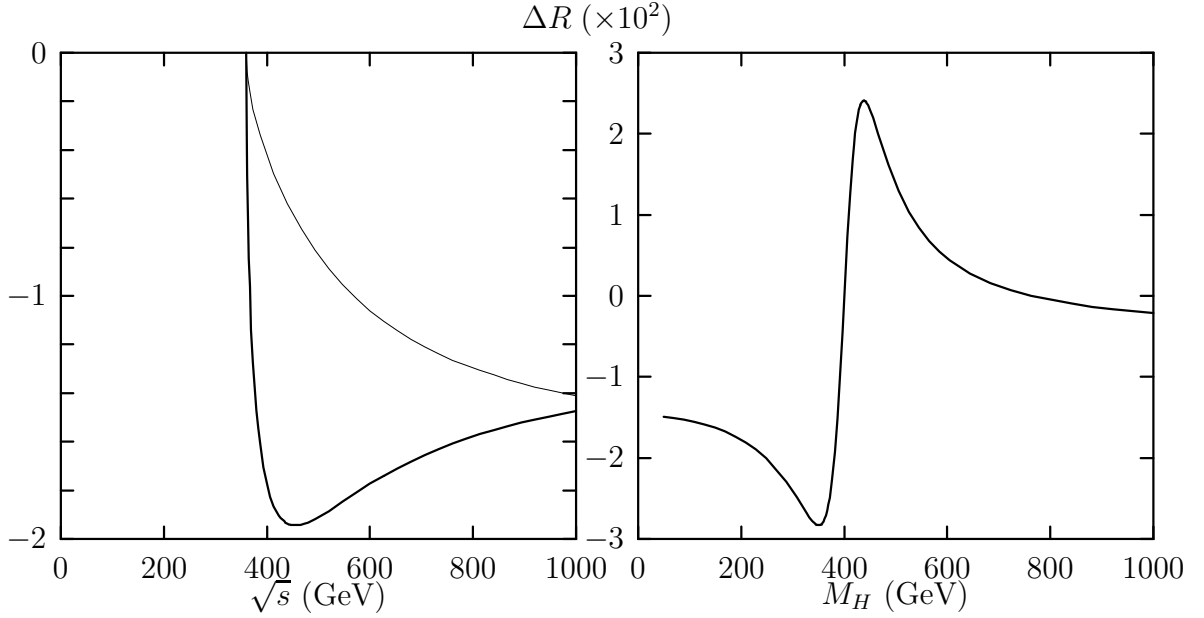

\section{\bf Visible $s-$channel Higgs contribution}
\label{sec:schannel}

Further, from Fig. \ref{fig:diagrams} it is clear that the $s-$channel Higgs 
contribution in the heavy-lepton collider would be significant, while in the 
electron collider is invisible due to the tiny electron mass \cite{muon}. 
This is the most important point, for example in the muon collider 
as pointed out in \cite{gunion}. This simple fact leads the author to 
combine $e^+ \, e^- \rightarrow q \, \bar{q}$ and 
${l_h}^+ \, {l_h}^- \rightarrow q \, \bar{q}$ processes at same 
center-of-mass energy to extract the $s-$channel Higgs contribution 
($l_h$ denotes arbitrary heavy-lepton and $q = b, t$), which is 
expected to be significant. This task can be done simply by set 
off the process with initial state $l_h$ againts the other one with 
initial state $e$, that is 
\begin{equation}
	\Delta \sigma \equiv \sigma^{l_h} - \sigma^e \; .
	\label{eq:deltasigma}
\end{equation}
Then, all of terms in Eq. (\ref{eq:msquare}) which are not multiplicated 
by ${m_l}^2$ will be exactly canceled out. Despite of another non 
${m_l}^2$ multiplicated terms from vector bosons mediated diagrams are 
remained, the contributions are expected to be comparable or dominated 
by the Higgs' one. Hence the $s-$channel Higgs contribution would be more 
visible. Corresponding with Eq. (\ref{eq:deltasigma}), the FB asymmetry, 
total CS and ratio $R$ become 
\begin{eqnarray}
	\Delta A_{\rm FB} & = & \Delta \sigma_F - \Delta \sigma_B \; , 
	\label{eq:deltafba} \\
	\Delta \sigma_T & = & \Delta \sigma_F + \Delta \sigma_B \; ,
	\label{eq:deltatcs} \\
	\Delta R & = & \frac{\Delta A_{\rm FB}}{\Delta \sigma_T} \; .
	\label{eq:deltar}
\end{eqnarray}
These equations are the main points in the paper, and will be analysed 
further. The author points out that Eqs. (\ref{eq:deltafba}) $\sim$ 
(\ref{eq:deltar}) are sensitive to the $s-$channel Higgs contribution 
as shown below. In other words, by using the data from the electron 
and heavy-lepton collider at same center-of-mass energy, considering the 
FB asymmetry and total CS in both colliders should be a reliable way to 
confirm the existance of $s$-channel Higgs.

In the numerical calculations, the parameters have been put as \cite{pdg}, 
$m_e = 0.51$ (MeV), $m_\mu = 105.66$ (MeV), $m_b = 4.3$ (GeV), 
$m_t = 180$ (GeV), $M_Z = 91.19$ (GeV), $M_W = 80.33$ (GeV), 
$\Gamma_Z = 2.49$ (GeV) and $\sin^2 \theta_W = 0.231$. For the 
Higgs decay width, approximately only the decays to fermionic 
final states except top quark, within tree-level SM are 
considered here \cite{higgswidth}. Note that the higher order corrections 
or more complete results should be seen in some references \cite{hwloop}. 
However, this rough approximation for the Higgs decay width is not so 
important for our interest in the present letter. 

Because the recent study on the heavy-lepton collider is focused 
in the muon collider, let us consider only the muon case in the 
present letter. 
The results for the electron vs muon collider case are presented 
in Figs. \ref{fig:tb} $\sim$ \ref{fig:ratio}. As mentioned before, 
the physical region is from $\sqrt{s} \geq 10$ (GeV) for bottom 
and $\sqrt{s} \geq 360$ (GeV) for top quark final states. 
In Fig. \ref{fig:tb}, the ratio $\Delta R$ for bottom and top quark final 
states is presented. It seems that observing the bottom quark final 
state process is better due to its higher rate. However, the author 
has checked that in the bottom quark final state case, the $s$-channel 
Higgs contributions are tiny and negligible, i.e. no visible discrepancy 
when the $s$-channel Higgs are included or not. Hence further analysis 
will be done only for top quark final state case. The (unnormalized) 
FB asymmetry and total CS for top quark final state are then presented 
in Fig. \ref{fig:tcsfba} with varying $\sqrt{s}$. In Fig. \ref{fig:ratio}, 
the Higgs mass and center-of-mass energy are fixed to be  $M_H = 200$ 
(GeV) for the left figure, and on the other hand $\sqrt{s} = 400$ (GeV) 
for the right one.

\section{\bf Discussion}
\label{sec:discussion}

From Fig. \ref{fig:tb}, the deviation due to different flavor of quark 
final states is significant, and is going to be larger for larger mass 
difference between them. It have also been checked that the deviations 
will be larger as considering electron vs heavier-lepton collider, 
like tauon collider. Especially in the very heavy quark final state case, 
like top quark, the deviation in all quantities seems large,  
as depicted in Figs. \ref{fig:tcsfba} and \ref{fig:ratio}. It can 
be concluded that in general, FB asymmetry is more sensitive than 
the total CS for any quark final states in electron vs any heavier-lepton 
colliders. The reason is, in the FB asymmetry most of the vector-bosons 
contributions are canceled out, while in the total CS the contributions 
are still remained. This result leads to the importance of $\Delta R$ 
which is newly defined in the previous section. However, the visibility 
will be reduced when one considers any light quark final states which 
are lighter than bottom quark, even for FB asymmetry. 

Finally, a rough estimation for required luminosity to observe $\Delta R$ 
can be given under assumptions that $\sqrt{s} \geq 400$ (GeV) and the 
event numbers is $10$ at one year running of machines. For the most 
optimistic case, the integrated luminosity is required to be larger 
than $100$ (fb$)^{-1}$. 
Remark that $\sqrt{s} = 400$ (GeV) with the integrated luminosity 
$\geq$ few (fb$)^{-1}$ is considered to be available in the 
muon collider (First Muon Collider, FMC) \cite{muon}. 
However, any analysis of the possible background effects in the 
lepton collider under consideration (both electron and muon colliders 
in the current case) must be studied further. The details of 
study for the background effects will be published elsewhere. 

Lastly, although the idea that there might be two lepton 
(with different flavor) colliders with the same center-of-mass 
energy seems farfetched, the current paper points out and shows an 
example that there may be any interesting physics by constructing a 
heavy-lepton collider with same center-of-mass energy as the 
present electron collider.
The author also hopes the study will encourage the interest in 
the possibility of constructing the heavy-lepton collider like the 
muon collider which should complement the present electron collider 
to examine the SM as well as open a window for new physics.

\bigskip
\noindent
{\Large \bf Acknowledgements}

The author thanks to Y. Kiyo for useful discussion in the beginning of 
this work and to T. Nasuno and I. Watanabe during the work. The author 
also expresses his gratitude to the Japanese Government for financially 
supporting his work under Monbusho Fellowship.


\begin{thebibliography}{1}
	\bibitem{electron}
		H. Murayama and M. E. Peskin, 
		to appear in {\it Ann. Rev. Nucl. Part. Sci.}.
	\bibitem{muon}
		{\it The Proceedings of 1st International Conference on 
		Physics Potential and Development of $\mu^+ \mu^-$ Colliders},
		Napa - California (1992),  
		{\it Nucl. Instru. and Meth.} {\bf A350} (1994); \\
		{\it The Proceedings of 2nd International Conference on 
		Physics Potential and Development of $\mu^+ \mu^-$ Colliders}, 
		Sausalito - California (1994), edited by D. Cline, 
		{\it American Institute of Physics Conference Proceeding}
		{\bf 352}; \\
		{\it The Proceedings of 3rd International Conference on 
		Physics Potential and Development of $\mu^+ \mu^-$ Colliders}, 
		San Fransisco - California (1995); \\
		J. F. Gunion, 
		talk presented at 
		{\it European Conference on High Energy Physics} in Brussels 
		(1995); \\
		S. A. Bogacz and D. B. Cline, 
		{\it Intl. Jour. Mod. Phys.} {\bf A11} (1996) 2613. 
	\bibitem{gunion}
		J. F. Gunion, 
		{\it Proceedings of the Rencontres de Physique de la 
		Valle d'Aoste} (1996), hep-ph/9605396; \\
		V. Barger, M. S. Berger, J. F. Gunion and T. Han,  
		hep-ph/9604334; \\
		V. Barger, M. S. Berger, J. F. Gunion and T. Han, 
		hep-ph/9602415. 
        \bibitem{pdg}
                Particle Data Group, 
                {\it Phys. Rev.} {\bf D54} (1996) 1.
	\bibitem{higgswidth}
		J. Ellis, M. K. Gaillard and D. V. Nanopoulos, 
		{\it Nucl. Phys.} {\bf B106} (1976) 292.
	\bibitem{hwloop}
		J. Fleischer and F. Jegerlehner, 
		{\it Phys. Rev.} {\bf D23} (1981) 2001; \\
		S. Dawson and S. Willenbrock, 
		{\it Phys. Lett.} {\bf B211} (1988) 200; \\
		E. Braaten and J. P. Leveille, 
		{\it Phys. Rev.} {\bf D22} (1980) 715; \\
		N. Sakai, 
		{\it Phys. Rev.} {\bf D22} (1980) 2220; \\
		T. Inami and T. Kubota, 
		{\it Nucl. Phys.} {\bf B179} (1981) 171.
\end{thebibliography}
\end{document}